\documentstyle[preprint,aps]{revtex}
\begin{document}
\draft
\title{Electrostatic ion perturbations in unmagnetized\\ plasma shear flow}
\author{A. D. Rogava}
\address{International Centre for Theoretical Physics, Trieste, Italy, and 
Department of Theoretical Astrophysics,  Abastumani  Astrophysical 
Observatory, Tbilisi 380060, Republic of Georgia}
\author{G. D. Chagelishvili}
\address{Abastumani  Astrophysical Observatory, Tbilisi, Republic of 
Georgia, and Department of Cosmogeophysics, Space Research Institute, 
Moscow 117810, Russia}
\author{V. I. Berezhiani}
\address{International Centre for Theoretical Physics, Trieste, Italy, and 
Department of Plasma Physics,  Institute of Physics, Tbilisi, Republic 
of Georgia}

\date{\today}
\maketitle
\begin{abstract}
The electrostatic perturbations in an unmagnetized, 
non-isothermal ($T_i{\ll}T_e$)
electron-ion plasma shear flow are considered. New physical effects, arising 
due to the non-normality of linear dynamics, are described. It is shown that 
the velocity shear induces the extraction of the mean flow energy by the
acoustic perturbations (ion-sound waves). The influence of the medium 
dispersion, rising due to the violation of quasineutrality for perturbations, 
is examined. It is shown that in the course of the evolution ion-sound waves 
turn into ion plasma oscillations. New class of 
nonperiodic, electrostatic perturbations (with vortical motion of ion
component), characterized by the intense energy exchange with the mean flow, 
is also described.

\end{abstract}

\pacs{52.30.-q, 52.35.Bj, 94.30.Tz, 97.60.Gb}

Classical stability theory of continuous media motion (normal mode approach) 
has been victorious in explaining how different kinds of shear flows become 
unstable. However, in some, quite simple and important kinds of {\it parallel
free shear flows} (e.g., plane Couette and Poiseuille, or pipe Poiseuille 
flows) the approach has serious problems, evoked by the not self-adjoint 
character of the governing equations [1--3]. That is why the predictions of 
the traditional stability approach fail to match the results of most 
experiments with these kinds of flows [3]. 

An alternative approach to the problem is that due to Kelvin [4], 
which implicates the change of independent variables from a laboratory to a
moving frame and study of the nonexponential temporal evolution of
{\it spatial Fourier harmonics} (SFH). The method is operative for any mean 
velocity profiles but it is most manageable to ones that are linear, or 
peacewise linear [1,5]. Effectiveness of the method was repeatedly proved, 
when it helped to obtain the bunch of unlooked-for results on the dynamics 
of the perturbations in hydrodynamical [6--12] and hydromagnetic [13--15] 
shear flows.

In [12] authors have considered the evolution of two-dimensional (2D) SFH 
in a compressible, plane hydrodynamical Couette flow.
Their analysis, which involved the nonmodal approach, brought them to
the discovery of the new mechanism of the energy exchange between 
the mean flow and sound-type perturbations. In particular, it was 
shown that energy of SFH may grow linearly in time---perturbations extract 
the energy from the mean shear flow. This process appears to be quite
universal and one should expect that it may be also influential in the wide
variety of continuous media with the analogous kinematics. 

In this {\it report} we shall examine the case of electron-ion plasma
shear flow and show that the process of velocity shear induced energy 
transfer from the mean flow to the collective modes exists and can be quite 
efficient in this case too. Moreover, as we shall see, the peculiarity of the 
plasma state of the medium plays its distinctive role and leads to the whole
group of interesting new effects. 

As it is well-known, in the
collisionless unmagnetized plasma with $T_i{\ll}T_e$ (where $T_e$ and $T_i$
are electron and ion temperatures respectively) exists the weakly damped 
low-frequency, longitudinal, and therefore electrostatic, ion mode. When
its wave-length ${\lambda}{\sim}1/k$ ($k$ is wave number) greatly exceeds 
electron Debye length (${\lambda}{\gg}{\lambda}_{De}{\equiv}(T_e/4{\pi}e^2
n_0)^{1/2}$, then this low-frequency mode represents the {\it ion-sound wave} 
with constant phase 
velocity (`ion acoustic' speed): $C_{s}{\equiv}(T_e/m_i)^{1/2}$. However, the 
velocity shear induces the "linear drift" of SFH (the process 
well-acknowledged in the literature [6,12,13] cultivating the nonmodal 
approach) which, mathematically, is
exposed in the temporal variation of the wave number vector ${\bf k}(t)$. 
It means that the influence of the {\it dispersion} of ion mode, arising as 
the result of the violation of quasineutrality for the perturbations, which, 
in other words, is related to the finiteness of $k{\lambda}_{De}$, should be 
taken into due account. The reason is simple: if initially 
$|k{\lambda}_{De}|{\ll}1$ the linear drift process will, eventually, transfer
the SFH to the region of ${\bf k}-space$, where the latter condition does not 
hold. In physical terms it means that that under certain
circumstances the ion sound waves, drawing energy from the mean shear flow,
subsequently turn into the ion plasma oscillations. The latter collective
mode is weakly damped if ${\lambda}{\gg}{\lambda}_{Di}
{\equiv}(T_i/4{\pi}e^2n_0)^{1/2}$. 

Velocity shear of the mean flow induces, also, excitation of the completely
new kind of non-periodic, electrostatic  perturbations with vortical motion
of the plasma ion component. These perturbations are able to the effective 
exchange of their energy with the mean flow and under certain conditions may 
play the {\it dominant} role in the behavior of the plasma flow.

At the end of this report we shall point out on the possible relevance of 
above noted remarkable processes to the problems of excitation of 
electrostatic waves, resonant acceleration of particles and onset 
of a turbulence in non magnetized electron-ion plasmas.

Let us consider electrostatic oscillation modes in the shear flow of 
electron-ion nonmagnetized plasma. The mean velocity ${\bf U}_0{\equiv}
(Ay,0,0)$ is directed along the X axis and has a linear shear along the 
$Y$ axis. Without the loss of generality we can assume that the constant
parameter $A$ is positive. Equilibrium electric field ${\bf E}$ is zero. 
Taking approval of 
the electrostatic nature of oscillations we can write the electric field
perturbation as the gradient of the electric potential: 
${\bf E}=-{\nabla}{\phi}$. The plasma in its equilibrium 
state is assumed to be homogeneous and quasineutral---equilibrium number 
density of electrons $n_{e0}$ is equal to the one of ions $n_{i0}$ and is 
constant: $n_{e0}=n_{i0}=n_0=const$. 

When the plasma is properly rarefied, ion and electron components interact 
weakly with each other and may, thus, have very different temperatures.
In particular, we shall assume that temperature of electrons is homogeneous
$T_e=const$ and for the sake of simplicity take that $T_{i}=0$. 
The effects, which may be evoked by the finiteness of $T_i$ will be discussed 
briefly at the end of the report. 

The usual approximation for electrostatic low-frequency waves implies 
that the electron number density can be described by the Boltzmann 
distribution [16]:
$$
n_e=n_{0}exp(e{\phi}/T_e){\approx}n_0(1+e{\phi}/T_e), \eqno(1)
$$

Let us, now, decompose the instantaneous values of all physical variables 
onto their mean and perturbed components: ${\bf V}_i={\bf U}_0+{\bf u}$,  
$n_{e,i}=n_0+n_{e,i}^{\prime}$. The basic system of linearized equations 
for ions, describing the evolution of small-scale, 3D perturbations in 
this flow, takes the form:
$$
[{\partial}_t+Ay{\partial}_z]n^{\prime}_i+n_0[{\partial}_{x}u_x+
{\partial}_{y}u_y+{\partial}_{z}u_z]=0, \eqno(2)
$$
$$
[{\partial}_t+Ay{\partial}_x]{\bf u}+({\bf u},{\nabla}){\bf U}_0
=-{e \over m_i}{\nabla}{\phi}, \eqno(3)
$$
$$
{\left[{\partial}^2_{x}+{\partial}^2_{y}+{\partial}^2_{z}\right]}
{\phi}=4{\pi}e(n^{\prime}_e-n^{\prime}_i). \eqno(4)
$$
where $({\bf u},{\nabla}){\bf U}_0$ is a vector, which has only
one, x-th, nonzero component equal to $Au_y$.

Let us make the following substitution of variables: $x_{1}=x - Ayt;~~
y_{1}=y;~~z_{1}=z;~~t_{1}=t$ and perform the Fourier analysis of (2--4), 
expanding unknown functions with respect to spatial coordinates: 
$$
F=\int{dk_{x_1}dk_{y_1}dk_{z_1}{\hat F}(k_{x_1},k_{y_1},k_{z_1},t_1)
exp[i(k_{x_1}x_1+k_{y_1}y_1+k_{z_1}z_1)]}, \eqno(5)
$$
where under $F$ we imply all physical variables appearing in the problem. 
Instead of equations (2) and (3) we get (for details, see [9,13]) the 
following set of first order, ordinary differential equations (ODE's) for  
SFH of these variables:
$$
N_i^{(1)}=v_x+{\beta}(\tau)v_y+{\gamma}v_z, \eqno(6)
$$
$$
v_x^{(1)}=-Rv_y-{\Phi}, \eqno(7a)
$$
$$
v_y^{(1)}=-{\beta}(\tau){\Phi}, \eqno(7b)
$$
$$
v_z^{(1)}=-{\gamma}{\Phi}; \eqno(7c)
$$
complemented by the algebraic relations, arising from (1) and (4):
$$
N_e={\Phi}, \eqno(8)
$$
$$
N_i={\left[1+{\xi}^2{\cal K}^2\right]}N_e, \eqno(9)
$$
where, hereafter, $F^{(n)}$ will denote the $n$-th order time derivative
of $F$.

In (6--9) we have introduced the following dimensionless notations:
$R{\equiv}A/C_{s}k_{x_1}$, ${\tau}{\equiv}C_{s}k_{x_1}t_1$, 
${\beta}_0{\equiv}k_{y_1}/k_{x_1}$, ${\beta}{\equiv}{\beta}_0-R{\tau}$, 
${\gamma}{\equiv}k_{z_1}/k_{x_1}$, ${\bf v}{\equiv}{\bf {\hat u}}/C_{s}$, 
${\xi}{\equiv}k_{x_1}C_{s}/{\omega}_{p}={\lambda}_{De}k_{x_1}$, 
${\Phi}{\equiv}ie{\hat {\phi}}/T_e$,
$N_{i,e}{\equiv}i{\hat n}^{\prime}_{i,e}/n_0$. 
Here ${\omega}_{p}{\equiv}(4{\pi}n_0e^2/m_i)^{1/2}$ is 
the ion plasma frequency. Note that ${\cal K}{\equiv}|k(t_1)|/k_{x_1}
=(1+{\beta}^2(\tau)+{\gamma}^2)^{1/2}$ is the modulus of the dimensionless 
and time-dependent wave number vector of SFH.

Note that the time-dependence of ${\beta}(\tau)$ and ${\cal K}(\tau)$ is
provoked by the temporal variability of the SFH wave number component along 
the flow shear:
$$
k_y(\tau){\equiv}k_{y_1}-Rk_{x_1}{\tau}=k_{x_1}{\beta}(\tau). \eqno(10)
$$
This process of the "linear drift" in the {\bf k}--space [13] has the 
crucial role for the temporal evolution of SFH. 

The total energy of perturbations consists of the kinetic energy of
ions, compressional energy associated with electrons (ions does not 
contribute, because their temperature is zero) and the energy of electric 
field. The phenomenological expression for the spectral density of the 
total energy is equal to:
$$
{\cal E}{\equiv}{1 \over 2}{\biggr{\{}|v_x|^2+|v_y|^2+|v_z|^2+[1+{\xi}^2
{\cal K}^2]|{\Phi}|^2{\biggl{\}}}}. \eqno(11)
$$
The first order ODE for ${\cal E}(\tau)$, derived from (6--9) is:
$$
{\cal E}^{(1)}=-R{\biggr[}v_x v_y+
2{\xi}^2{\beta}(\tau){\Phi}^2{\biggl]}, \eqno(12)
$$
and in the "no-shear" ($R=0$) limit it is the conserved quantity.

It is easy to verify that variables appearing in Eq. (6--9) obey the
following remarkable {\it algebraic} relation:
$$
(1+{\gamma}^2)v_y-{\beta}(\tau)(v_x+{\gamma}v_z)=RN_i+const. \eqno(13)
$$

By means of this relation, taking a derivative of eq. (6) and re-arranging
terms, we can get the following explicit second order ODE for the function
${\Psi}{\equiv}N_i/{\cal K}$:
$$
{\Psi}^{(2)}+{\Omega}^2(\tau){\Psi}=const{\times}f(\tau), \eqno(14)
$$
where
$$
{\Omega}^2(\tau){\equiv}{{{\cal K}^2}\over
{1+{\xi}^2{\cal K}^2}}+{{3R^2(1+{\gamma}^2)}\over
{{\cal K}^4}}, \eqno(15a)
$$
$$
f(\tau){\equiv}-2R/{\cal K}^3. \eqno(15b)
$$

It should be noted that the analogous, but more simple, kind of inhomogeneous 
second-order ODE was derived and analyzed in [12].
The general solution of (14) is the sum of its {\it special} solution and the 
{\it general} solution of the corresponding homogeneous ($const=0$) 
equation: ${\Psi}={\Psi}_h+{\Psi}_s$.  When ${\Omega}(\tau)$ depends on 
$\tau$ adiabatically, implying [12,17]:
$$
|{{\Omega}(\tau)}^{(1)}|{\ll}{\Omega}^2(\tau), \eqno(16)
$$
then the homogeneous equation can be solved approximately.

For the flows with $R{\ll}1$ the condition (16) holds for the wide 
range of possible values of $|{\beta}(\tau)|$. In other words,
since the temporal variability of $|{\beta}(\tau)|$ is determined by the
"linear drift" of SFH, (16) is valid at all stages of the evolution of
the SFH. When the condition (16) holds, the approximate expression for 
${\Psi}_h$ may be written in the following way:
$$
{\Psi}_h(\tau){\approx}{C \over{\sqrt{{\Omega}(\tau)}}}exp[i({\varphi}(\tau)+
{\varphi}_0)], \eqno(17)
$$
where ${\varphi}(\tau)={\int}_0^{\tau}{\Omega}({\tau}^{'})d{\tau}^{'}$. 

As regards the special solution of inhomogeneous equation (14), it can
also be derived owed to the smallness of the $R$ parameter. In particular,
the solution may be expressed by the following series [18]:
$$
{\Psi}_s(\tau)=const{\times}\sum_{n=0}^{\infty}R^{2n}y_n(\tau), \eqno(18a)
$$
$$
y_0(\tau)=f(\tau)/{\Omega}^2(\tau), \eqno (18b)
$$
$$
y_n(\tau)=-{1 \over {\Omega}^2(\tau)}{{{\partial}^2y_{n-1}}
\over {{\partial}{\beta}^2}}. \eqno (18c)
$$

Since $R{\ll}1$, the terms with higher powers of $R$ are
negligible and the {\it full} approximate solution of the 
inhomogeneous equation (14) may be written explicitly as:
$$
{\Psi}={\Psi}_h+{\Psi}_s{\approx}{C \over {\sqrt{{\Omega}(\tau)}}}
exp[i({\varphi}(\tau)+{\varphi}_0)]+{{const{\times}f(\tau)}
\over{{\Omega}^2(\tau)}}. \eqno(19)
$$

When $C/const{\ll}1$ the SFH may be treated as mainly incompressible and
vortical perturbation, while when $C/const{\gg}1$ it is mainly
of the sound-type.

Having in hands the solution of (14) we are, certainly, able to find
all other variables of the problem. They may be calculated by ${\Psi}$ and 
${\Psi}^{(1)}$ in the following way:
$$
N_i={\cal K}{\Psi}, \eqno(20)
$$
$$
N_e={\cal K}{\Psi}/(1+{\xi}^2{\cal K}^2), \eqno(21)
$$
$$
v_y={1 \over {{\cal K}^3}}{\biggl[}{\beta}{\cal K}^2{\Psi}^{(1)}
+R(1+{\gamma}^2)
{\Psi}+const{\times}{\cal K}{\biggr]}, \eqno(22)
$$
$$
v_x+{\gamma}v_z={1 \over {{\cal K}^3}}{\biggl[}(1+{\gamma}^2){\cal K}^2
{\Psi}^{(1)}-R{\beta}(1+{\gamma}^2+{\cal K}^2){\Psi}
-const{\times}{\beta}{\cal K}{\biggr]}, \eqno(23)
$$

Note that (20--23) are exact expressions, valid for {\it arbitrary} values
of the shear parameter $R$. For large enough values of $R$, when adiabatic
solution (19) is no longer valid, Eqs. (20--23) may still be used for 
reproduction of the variables through ${\Psi}$ and ${\Psi}^{(1)}$, obtained, 
this time, by the direct numerical solution of equation (14). 

Below we shall focus our attention on the behavior of 2D perturbations in 
the $XOY$ plane (${\gamma}=0$), when the shear parameter $R{\ll}1$. This 
case admits simple analytical examination and exposes soundly the qualitative
novelty of the problem. Using expression (11) for the spectral energy 
density ${\cal E}(\tau)$ and (20--23), we get for ${\cal E}(\tau)$ the 
following simple expression:
$$
{\cal E}(\tau){\simeq}{1 \over 2}{\left[C^2{\Omega}(\tau)+
{\left({const \over {\cal K}(\tau)}\right)^2}\right]}. \eqno(24)
$$

The spatial characteristics of the SFH ($k_{x_1}$, $k_y(\tau)$) and the value 
of the shear parameter $R$ manage the evolution of the frequency of 
oscillations  and the actual intensity of the energy exchange between the
SFH and the background flow. In particular, temporal variability of these 
processes are essentially induced by the "linear drift" of SFH in the 
{\bf k}-space [13,12].

For the sound-type ($C{\ne}0$, $const=0$) perturbations, as it is evident from
(15a), the frequency of oscillations varies with the variation of 
${\cal K}(\tau)$. Originally, at moderate values of ${\cal K}(\tau)$, due
to the smallness of $\xi$, the oscillation mode may be treated as the 
ion-sound wave (${\Omega}(\tau){\sim}{\cal K}^2(\tau)$). Afterwards, when
${\cal K}(\tau)$ reaches large enough values, the dispersive influence of
the denominator in the first term of (15a) becomes more and more imperative 
and when ${\xi}^2{\cal K}^2(\tau){\gg}1$ the frequency ${\Omega}(\tau)$
already exhibits ion plasma oscillations. Following, according to (24), the 
evolution of energy  of this mode (${\cal E}(\tau){\sim}{\Omega}(\tau)$), we
find that  
initially, for ${\beta}_0>0$, at $0<{\tau}<{\tau}_*{\equiv}|{\beta}_0|/R$, 
the energy decreases and reaches its minimum at ${\tau}={\tau}_*$. 
A while later, it begins to increase 
at ${\tau}_*<{\tau}<{\infty}$, when the SFH
"emerges" into the area of ${\bf k}$-space in which $k_y(\tau)k_{x_1}<0$
(the "growth area" for the sound-type perturbations [12]). If the SFH is in 
the "growth area" from the beginning (${\beta}_0<0$), its energy increases 
monotonically. When ${\xi}{\cal K}(\tau){\ge}1$ the rate of the energy 
increase becomes less and less and the energy asymptotically tends to the 
constant value. 

When  $C=0$ and $const{\ne}0$ the SFH may be treated as mainly incompressible 
and vortical perturbations. In this case ${\Psi}{\simeq}const{\times}f(\tau)/
{\Omega}^2(\tau)$, while $v_y{\simeq}const/{\cal K}^2(\tau)$ and 
$v_x{\simeq}-const{\times}{\beta}/{\cal K}^2(\tau)$. 
The spectral energy of SFH 
varies as ${\cal E}(\tau){\simeq}{\cal K}^{-2}(\tau)$ and reduces to the  
well-known expression, describing the "transient" growth of the energy
of SFH [6,7,13]. Transient increase of the energy takes place if initially 
$k_{y_1}/k_{x_1}>0$ (${\beta}_0>0$) and occurs nearby the 
${\tau}_{*}{\equiv}{\beta}_0/R$ moment of time, when ${\beta}(\tau)$ tends to
zero and ${\cal K}(\tau)$ attains its minimum value. Such is the behavior of
2D vortical perturbations. 
One should expect that the evolution of three-dimensional 
(3D) vortical perturbations should be similar to the behavior of the
analogous structures in incompressible, inviscid fluids, extensively 
studied in [9].
As it was shown in [9], the energy of 3D vortical perturbations grows also 
nonexponentially, but unlike transiently growing 2D perturbations, the 
energy of 3D perturbations saturates, attaining in asimptotics some constant 
value.  

Above noted similarity of the behavior of vortical perturbations with the
same process in usual fluids [12], holds only in the low-$R$ ($R{\ll}1$)
range. For larger values of the $R$ parameter ($R{\simeq}1$) one should
expect notable differences between the behavior of vortical perturbations 
in neutral fluids and electron-ion plasma.

Certainly, in the general case ($C{\simeq}const$), the "vortical" 
and the "sound-type" evolution of  perturbations are superimposed on 
one another.

Thus, we see that the dispersive nature of plasma, arising due to the 
violation of the quasineutrality for density perturbations, leads to the
smooth transition of ion-sound waves, amplified by the process of a shear 
flow energy extraction, into the ion plasma oscillations. As regards to the
another type of dispersion properties of ion oscillations, related to the
finiteness of {\it ion} temperature (and, for simplicity, neglected in the
above consideration) they should lead to the transfer of the energy to the
shorter wave length region. The characteristic phase velocity of 
perturbations in the course of the "linear drift" of SFH, tends to 
ion-thermal velocity. Due to the resonant interaction with ions these 
perturbations will be damped (Landau damping) transferring their energy to
the particles and accelerating them. However, we should notice that such 
channel of dissipation and collisionless heating of plasma ion component is
relevant for  perturbations with characteristic time scale smaller, than
the time scale for electron-ion collisions. The energy can also become one 
of the main sources for the onset of plasma turbulence in such flows.
 
Surely, the problem under consideration is quite complex. The comprehensive
analysis of all possible regimes is beyond the scope of this  report
and will be published elsewhere. 
\vskip 1cm
\centerline{\large \bf Acknowledgements}
\vskip 0.5cm
ADR's and GDC's research was supported, in part, by International Science 
Foundation (ISF) long-term research grant RVO 300. VIB's research was 
supported, in part, by International Science Foundation (ISF) long-term 
research grant KZ3200. ADR's visit to ICTP was supported, in part, by the 
Committee for Science and Technology of Republic of Georgia.


\begin{references}
\item 
W.O. Criminale and P.G. Drazin Stud. Appl. Maths., {\bf83}, 123 (1990).
\item 
S.C. Reddy and D.S. Henningson, J. Fluid Mech. {\bf252}, 209 (1993).
\item
L.N. Trefethen, A.E. Trefethen, S.C. Reddy, and T.A. Driscoll,
Science {\bf261}, 578 (1993).
\item 
Lord Kelvin (W. Thomson), Phil. Mag. {\bf24}, Ser. {\bf5}, 188 (1887).
\item
W.O. Criminale, T.L. Jackson, and D.G. Lasseigne, J. Fluid. Mech.
{\bf294}, 283 (1995).
\item
P. Marcus, W.H. Press, J.Fluid Mech. {\bf 79}, 525 (1977).
\item
A.D.D. Craik, W.O.Criminale, Pros. R. Soc. Lond. A{\bf 406}, 13
(1986).
\item
J.G. Lominadze, G.D. Chagelishvili, and R.A. Chanishvili,
Pis'ma Astron. Zh. {\bf 14}, 856 (1988) [Sov. Astron.
Lett. {\bf 14}, 364 (1988)].
\item
G.D. Chagelishvili, R.G. Chanishvili, J.G.
Lominadze, and I.N. Segal, Proceedings of the fourth International
Conference on Plasma Physics and Controlled Nuclear Fusion, held in
Toki, Japan 17-20 November 1992 (ESA SP--351, 1993).
\item
K.M. Butler and B.F. Farrell, Phys. Fluids A {\bf 4}, 1637 (1992).
\item
L.H. Gustavsson, J.Fluid Mech. {\bf 224}, 241 (1991).
\item
G.D. Chagelishvili, A.D. Rogava, and I.N. Segal, Phys. Rev. (E)
{\bf 50}, 4283 (1994).
\item
G.D. Chagelishvili, T.S. Christov, R. G. Chanishvili, and J.G. Lominadze,
Phys. Rev. (E) {\bf 47}, 366 (1993).
\item
S.A. Balbus, and J.H. Hawley, Ap.J. {\bf 400}, 610 (1992).
\item
S.H. Lubow, and H.C. Spruit, Ap.J. {\bf 445}, 337 (1995).
\item
R.J. Goldston and P.H. Rutherford {\it Introduction to Plasma Physics}
(Institute of Physics Publishing, Bristol 1995).
\item
Zel'dovich Ya. B. and Mishkis A. D. {\it Elementi Prikladnoi
Matematiki} (Nauka, Moscow 1972) (in russian).
\item
Magnus K. 1976, {\it Schwingungen} (B. G. Teubner, Stuttgart).
\end{references}
\end{document}